\documentclass[aps,prb,showpacs,groupedaddress,twocolumn]{revtex4}
\usepackage{amsmath}
\usepackage{amssymb}
\usepackage{graphicx}
\usepackage{txfonts}
\usepackage{color}
\begin{document}

\title{Topological Anderson insulator phenomena}

\author{Yanxia Xing$^{1,2}$ Lei Zhang$^2$ and Jian Wang$^{2,\ast}$}
\address{
$^1$Department of Physics, Beijing Institute of Technology, Beijing
100081, China\\$^2$Department of Physics and the Center of
Theoretical and Computational Physics, The University of Hong Kong,
Pokfulam Road, Hong Kong, China}

\begin{abstract}
We study the nature of the disorder-induced quantized conductance,
i.e., the phenomena of topological Anderson insulator (TAI) induced
in HgTe/CdTe semiconductor quantum well. The disorder effect in
several different systems where anomalous Hall effect exist, is
numerically studied using the tight-binding Hamiltonian. It is found
that the TAI phenomena also occur in the modified Dirac model where
the quadratic corrections $k^2\sigma_z$ is included and
electron-hole symmetry is kept. It also occurs in the graphene
system with the next nearest-neighbor coupling and staggered
sublattice potential. Comparison between the localization lengths of
the 2D ribbon and 2D cylinder clearly reveals the topological nature
of this phenomena. Furthermore, analysis on the local current
density in anomalous quantum Hall systems where the TAI phenomena
can or can not arise reveals the nature of TAI phenomena: the bulk
state is killed drastically and only the robust edge state survives
in a moderate disorder. When the edge state is robust enough to
resist the strong disorder that can completely kills the bulk state,
TAI phenomena arise.
\end{abstract}

 \maketitle

\section{introduction}
It is known that the two-dimensional (2D) noninteracting system with
the quadratic dispersion relation is an Anderson
insulator.\cite{TAI}

XXX

In the presence of strong spin orbit coupling (SOC)\cite{soi,Qiao},
the external\cite{Qiao,ShengDN} or internal\cite{Haldane} magnetic
field, the metallic state is present. Due to the crossing of the
mobility edge\cite{Onoda}, a metal-insulator transition (MIT) occurs
at the critical disorder strength $w_c$ where the localization is
divergent. Recently, Li et al\cite{LiJian} find that the disorder
drive either a metallic state or an ordinary insulating state to the
topological insulator (it is so called topological Anderson
insulator) in the HgTe/CdTe quantum well, which has been numerically
confirmed by Jiang et el\cite{JiangHua}. Using the effective medium
theory, the mechanism of the TAI is explained as the crossing of a
band edge rather than a mobility edge by Groth et
al.\cite{Beenakker} Although there are many investigations focused
on the TAI, there are still some unanswered questions. For instance,
what leads to the band edge crossing, or, what's the nature of the
TAI phenomena (i.e., disorder induce quantized conductance)?
Furthermore, except for HgTe/CdTe quantum well, is there any other
systems that have TAI phenomena? What's the necessary condition to
generate TAI phenomena?

As we know, the edge states in topological
insulator\cite{topoInsulator} are ``helical" states, i.e., the
direction of propagation is tied to the electron spin rather than
electron charge, the edge states located in the opposite edges are
protected by the time reversal symmetry and each contribute $e^2/h$
to the conductance.
In principle, we can focus only on one edge state which is tied to
spin up or spin down to study topological insulator. The sub-system
related to each individual spin contribute to an individual
anomalous quantum Hall effect. In the following, we will deal with
the sub-system related to the spin up in topological insulator and
treat it as an individual anomalous quantum Hall effect.

In a 2D anomalous quantum Hall\cite{Qiao1,QAH} system, the
topological edge states connect the energetically separated
continuum of energy band, and only the unidirectional topological
edge state contribute to the conductance in the band gap. Due to the
topological stability of Chern numbers\cite{ChernNum} carried by the
extended states, the conductance remains quantized in the presence
of weak disorder. At strong disorders, the mobility edges is
crossed, and MIT\cite{Onoda,ShengDN} occurs. However, outside of the
gap, the bulk state and edge state may co-exist. Because the edge
state is robust for disorders, it could happen that the bulk state
is completely killed before the edge state is killed, which may lead
to the quantized conductance plateau. The quantized value is
determined by the number of the robust edge state. Taking into
account of edge state tied to both the spin up and spin down,
disorder induced TAI can be formed, which is confirmed by the
following calculation on the modified Dirac model. So, roughly
speaking, as long as the bulk states and the robust enough edge
state coexist in the system, disorder induced quantized conductance
can emerge.

In this paper, in order to effectively study the nature of TAI
phenomena, in addition to HgTe/CdTe quantum well,\cite{ZhangSC} we
first find two other different models in which TAI phenomena also
exists as in Ref.\onlinecite{LiJian}. The first model is a modified
Dirac model on a square lattice with the quadratic corrections
$k^2\sigma_z$ included and electron-hole symmetry kept. This model
is similar to HgTe/CdTe quantum well model except the e-h symmetry
is broken in HgTe/CdTe quantum well.\cite{ZhangSC} The second model
is a graphene model on a honeycomb lattice with the next
nearest-neighbor coupling and a staggered sublattice
potential\cite{Haldane}. In this model the e-h symmetry and
inversion symmetry are all broken, and the anti-directional
topological edge states tied to two opposite edges are
asymmetrically distributed. Our calculation shows that in the first
and second models, moderate disorder induces a transition from an
ordinary metallic state to a TAI with the quantized conductance of
$G_0=e^2/h$.
The wider the ribbon is, the smaller the fluctuation. In the
disorder induced quantized conductance regime, the localization
length of the ribbon structure is extremely long comparing to that
of the cylinder structure, which clearly reveals the topological
origin of the transition. When the disorder is very strong, mobility
edge crosses and MIT occurs. This process is related to the bulk
state. In order to vividly show how the transport electron is
scattered by the disorder, we calculate the individual local current
density from all channels including the edge state channel and bulk
state channel. In addition to the first and second model, we also
calculate local current density in a third model: the honeycomb
graphene system\cite{Qiao1} in which we consider Rashba spin-orbit
interaction, exchange energy and staggered sublattice potential. In
the third model, the disorder can't quantize the conductance but
just flatten it, which is different from the first and second model.
However, for all three models, we find that disorder kill all the
bulk current and induce edge current with uni-direction. It means
that in the process of transport, due to the topological nature, the
edge state is maintained all the time, no matter whether it is kept
in the edge channel or scattered into the bulk channels. So, in the
system where the edge state and bulk state coexist, in the moderate
disorder, the edge state is maintained while the bulk states are
killed, which leads to the flattened conductance. If the edge state
is robust enough to resist the disorder that is so strong that the
bulk state are completely killed, quantized conductance can be
formed.

The rest of the paper is organized as follows. In Sec. II, the
Hamiltonian of three model systems in the tight-binding
representation are introduced. The formalisms for calculating the
conductance and the local current density vector are then derived.
Sec. III gives numerical results along with some discussions.
Finally, a brief summary is presented in Sec. IV.

\section{models and formalism}
\subsection{three model Hamiltonian}
\begin{figure}
\includegraphics[bb=1mm 101mm 208mm 202mm,
width=8.5cm,totalheight=4.3cm, clip=]{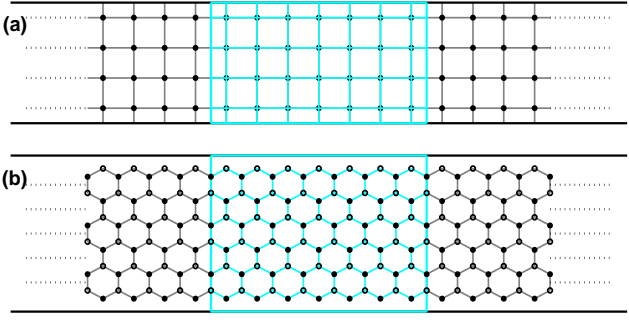} \caption{
(Color online) A schematic diagram of the infinite long ribbon in
square lattice for the first model [panel (a)] and the infinite long
ribbon in honeycomb lattice along the zigzag direction for the
second and third model [panel (b)]. The left lead and right lead
(the red area) in the p-region [green lattice region].}
\label{structure}
\end{figure}
The first model is  the modified Dirac model with a quadratic
corrections $k^2\sigma_z$, which has the form
\begin{eqnarray}
&&H_1 = \sum_{\bf k}[H_{\uparrow}({\bf k})+H_{\downarrow}({\bf
k})],~~
H_{\downarrow}({\bf k})=H^*_{\uparrow}(-{\bf k})   \nonumber \\
&&H_{\uparrow}({\bf k}) = A({\bf k}_x\sigma_x-{\bf
k}_y\sigma_y)+(m+B{\bf k}\cdot{\bf
k})\sigma_z+\epsilon(r)\sigma_0\label{Dirack}
\end{eqnarray}
where $\sigma_{x,y,z}$ are Pauli matrices presenting the pseudospin
formed by $s,p$ orbitals, In Eq.(\ref{Dirack}), the momentum $k$ is
a good quantum number for periodic systems. This model is similar to
the low-energy effective Hamiltonian of a HgTe/CrTe quantum
well\cite{ZhangSC} except that the e-h symmetry is kept here while
it is broken in HgTe/CrTe quantum well. Using substitution ${\bf
k}_x\rightarrow-i\partial_x$ and ${\bf k}_y\rightarrow-i\partial_y$
Eq.(\ref{Dirack}) can be transformed into the real space
representation.\cite{book} In the calculation we consider the
infinite long ribbon with finite width, hence the tight-binding
representation is convenient. The tight-binding Hamiltonian in
square lattice is given by:\cite{LiJian,JiangHua}
\begin{eqnarray}
H_{\uparrow} &=& \sum_{\bf i} d^{\dagger}_{\bf i}
\left(\epsilon_{\bf i}\sigma_0 +C\sigma_z\right) d_{\bf i}\nonumber
\\ &+&\sum_{\bf i} d_{\bf i}^{\dagger}
\left(t\sigma_z-i\frac{A}{2a}\sigma_x\right) d_{{\bf i}+{\bf a}_x}
+h.c.\nonumber
\\ &+&\sum_{\bf i} d_{\bf i}^{\dagger}
\left(t\sigma_z+i\frac{A}{2a}\sigma_y\right) d_{{\bf i}+{\bf
a}_y}+h.c. \label{Diracr}
\end{eqnarray}
where $\sigma_0$ is a unitary $2\times2$ matrix, $\epsilon_{\bf i}$
is a random on-site potential which is uniformly distributed in the
region $[w/2,w/2]$. ${\bf i}=({\bf i}_x , {\bf i}_y)$ is the index
of the discrete site of the system sketched in the Fig.1(a) on the
square lattice, ${\bf a}_x=[a,0]$ and ${\bf a}_y=[0,a]$ are the unit
vectors of the square lattice with the lattice constant $a$. $d_{\bf
i}=[d_{s,{\bf i}},d_{p,{\bf i}}]^T$ with $`T'$ denoting transpose
and $d_{s(p),{\bf i}}$ and $d_{s(p),{\bf i}}^{\dagger}$ are the
annihilation and creation operators for s(p) orbital at site ${\bf
i}$. Here $C=m-4t$ and $t=B/a^2$ denote the nearest neighbor
coupling strength. $\epsilon_{\bf i}$ in the first term of
Eq.(\ref{Diracr}) is the on-site random energy account for the
disorder. The second term in Eq.(\ref{Diracr}) is the corrected
linear Dirac term, in which the $\sigma_z$ involved terms are the
quadratic corrections to the Dirac Hamiltonian. The individual spin
up Hamiltonian $H_\uparrow$ and spin down Hamiltonian $H_\downarrow$
in Eq.(\ref{Diracr}) are time reversal symmetric to each other.
Since they are decoupled, we can deal with them individually. So we
shall focus only on spin-up Hamiltonian $H_\uparrow$ in the
following calculation.

The second model is proposed by Haldane that considers the graphene
with next-nearest neighbor coupling and staggered sublattice
potential whose Hamiltonian can be expressed as:\cite{Haldane}
\begin{eqnarray}
&&H_2({\bf k}) =
-t\sum_{i}\left[\epsilon(r)\sigma_0+t_o\sigma_z+\cos({\bf k}\cdot
{\bf a}_i)\sigma_x-\sin({\bf k}\cdot {\bf a}_i)\sigma_y\right]
\nonumber \\
&&+ 2|t_n|\left[\cos\phi\sum_i \cos({\bf k}\cdot {\bf
b}_i)\sigma_0-\sin\phi\sum_i \sin({\bf k}\cdot {\bf
b}_i)\sigma_z\right] \label{Haldanek}
\end{eqnarray}
where $\sigma_{x,y,z}$ are Pauli matrix denoting the pseudospin
formed by AB sublattice, three nearest neighbor unit vectors ${\bf
a}_{i=1,2,3}$ and next-nearest neighbor unit vectors ${\bf
b}_{i=1,2,3}$ are given by ${\bf a}_1=a(0,1)$, ${\bf
a}_2=a(-\sqrt{3}/2,-1/2)$, ${\bf a}_3=a(\sqrt{3}/2,-1/2)$, ${\bf
b}_1=b(1,0)$, ${\bf b}_2=b(-1/2,\sqrt{3}/2)$, ${\bf
b}_3=b(-1/2,-\sqrt{3}/2)$ with $a=0.142nm$ and $b=\sqrt{3}a$
denoting the distance between nearest neighbor sites (lattice
constant) and next-nearest neighbor sites respectively. In
Eq.(\ref{Haldanek}), $t_o$ is the staggered sublattice potential,
$t=2\hbar v_F/3a$ is the nearest neighbor coupling strength with
Fermi velocity $v_F=0.89\times10^6ms^{-1}$, the next-nearest
neighbor coupling $t_n=|t_n|e^{i\phi}$ where $|t_n|$ and $\phi$ are
the coupling strength and phase deduced from the effective internal
magnetic field $B(r)$ along $\vec z$ direction. Here $t_n$
eliminates the e-h symmetry of the energy bands as shown in Fig.2(b)
and $B(r)$ breaks the time-reversal invariance. Eq.(\ref{Haldanek})
is similar to Eq.(\ref{Dirack}), which includes all the
$\sigma_{x,y,z}$ terms that is $k$-dependent. However, they have
different symmetries, e.g., in Eq.(\ref{Haldanek}) both
time-reversal and e-h symmetries are broken. In the tight-binding
representation, Eq.(\ref{Haldanek}) can be expressed as:
\begin{eqnarray}
H_2 &=& \sum_{\bf i}d^{\dagger}_{\bf i}(t_o\sigma_z+\varepsilon_{\bf
i})d_{\bf i}+t_c\sum_{{\bf i},i}
(a^{\dagger}_{\bf i} b_{{\bf i}+{\bf a}_i}+h.c.)\nonumber \\
&+&|t_n|\sum_{{\bf i},i} \left[e^{i\phi}(a_{\bf i}^{\dagger} a_{{\bf
i}+{\bf b}_i}-b_{\bf i}^{\dagger} b_{{\bf i}+{\bf
b}_i})\right]+h.c.\label{Haldaner}
\end{eqnarray}
where ${\bf i}=({\bf i}_x , {\bf i}_y)$ is the index of the discrete
site of the honeycomb lattice which includes two sublattice A (open
circle) and B (filled circle) as sketched in the Fig.1(b). $\bf
\varepsilon_{\bf i}=diag(\epsilon_{\bf i},\epsilon_{{\bf i}+{\bf
a}_1}$) denotes the random potential induced by disorder, $d_{\bf
i}=[a_{\bf i},b_{{\bf i}+{\bf a}_1}]^T$, $a_{\bf i}(b_{\bf i})$ and
$a^\dagger_{\bf i}(b^\dagger_{\bf i})$ are the annihilation and
creation operators for sublattice A(B) at site ${\bf i}$.

In the third model, we consider a graphene sheet with Rashba
spin-orbit interaction $\lambda \vec{e}_z\cdot({\bf k}\times{\bf
s})$ as well as an exchange field.\cite{Qiao1}
In the tight-binding representation on the honeycomb lattice,
corresponding to three directions of translational symmetry, we have
${\bf k}_{\bf{i}=1,2,3}\rightarrow -i\partial_{\bf i}$. Projecting
them to the $x$ and $y$ direction, we can get ${\bf k}_x$ and ${\bf
k}_y$. Then, the Hamiltonian of the third graphene model including
Rashba spin-orbit coupling, the staggered sublattice potential and
the exchange term can be expressed in the tight-binding
representation in the following form:\cite{ShengL}
\begin{eqnarray}
H_3 &=& \sum_{\bf i}d^{\dagger}_{\bf i}(t_o\sigma_z\otimes
s_0+t_e\sigma_0\otimes s_z+\varepsilon_{\bf i})d_{\bf i} \nonumber
\\&+& \sum_{{\bf i},i}a^{\dagger}_{\bf
i}\left[t_cs_0+it_r\vec{e}_z\cdot({\bf s}\times\vec{\bf a}_i)\right]
b_{{\bf i}+{\bf a}_i}+h.c.\label{Qiaor}
\end{eqnarray}
where $s_0$ is the unitary matrix in spin space, ${\bf
s}=(s_x,s_y)$, $s_{x,y,z}$ are Pauli matrix denoting the real spin,
$d^\dagger_{\bf i}=[a^\dagger_{\bf i},b^\dagger_{{\bf i}+{\bf
a}_1}]^T$, $a^\dagger_{\bf i}=[a^\dagger_{\uparrow,{\bf
i}},a^\dagger_{\downarrow,{\bf i}}]$, $b^\dagger_{\bf
i}=[b^\dagger_{\uparrow,{\bf i}},b_{\downarrow,{\bf i}}]$ where
$a^\dagger_{\uparrow(\downarrow),{\bf i}}$ and
$b^\dagger_{\uparrow(\downarrow),{\bf i}}$ are the creation
operators for sublattice A and B for spin up(down) at site ${\bf
i}$. $\bf \varepsilon_{\bf i}=diag(\epsilon_{\bf
i}s_0,\epsilon_{{\bf i}+{\bf a}_1}s_0$) is the random potential
induced by disorder. Similar to the Hamiltonian of the second model,
$t_c$ is the nearest neighbor coupling and $t_o$ describes the
staggered sublattice potential. In Eq.(\ref{Qiaor}), $t_e$ is the
exchange energy which can be achieved by either a magnetic insulator
substrate or adsorbing transition metal atoms (e.g., iron, copper)
on graphene.\cite{Qiao1} Finally $t_r$ is the strength of Rashba
spin-orbit coupling that has been shown to be fairly
strong.\cite{Rashba}

\subsection{differential current density and conductance}

In the following, we consider two geometries: ribbon and cylinder.
Fig.1(a) and (b), respectively, depict the ribbon geometry on square
lattice and honeycomb lattice. In these systems the finite
scattering region [green (or gray in print) region], in which random
disorder is considered, is connected to the external reservoir
through semi-infinite lead. In the following, we will derive the
conductance and the current density in the scattering region.

For the general Hamiltonian $H=\sum_{\sigma{\bf i},\sigma'{\bf
j}}c^\dagger_{\sigma{\bf i}}{\bf H}_{\sigma{\bf i},\sigma'{\bf
j}}c_{\sigma'{\bf j}}$, local current flowing from site ${\bf i}$
with real spin or pseudospin $\sigma$ can be expressed
as:\cite{Jauho}
\begin{eqnarray}
{\bf J}_{\sigma,{\bf i}}(t) &=& -e\langle \dot{N}_{\sigma,{\bf
i}(t)}\rangle=\frac{ie}{\hbar}\langle c^\dagger_{\sigma,{\bf
i}}(t)c_{\sigma,{\bf i}}(t),{\bf H} \rangle
\nonumber \\
&=&\frac{e}{\hbar}\sum_{\sigma'{\bf j}}\left[{\bf G}^<_{\sigma{\bf
i},\sigma'{\bf j}}(t,t){\bf H}_{\sigma'{\bf j},\sigma{\bf i}}-{\bf
H}_{\sigma{\bf i},\sigma'{\bf j}}{\bf G}^<_{\sigma'{\bf
j},\sigma{\bf i}}(t,t)\right]
\nonumber \\
&=&\sum_{\sigma'{\bf j}}{\bf J}_{\sigma{\bf i},\sigma'{\bf j}}
\end{eqnarray}
where $e$ is the electron charge, ${\bf G}^<_{s{\bf i},s'{\bf
j}}=i\langle c^\dagger_{s'{\bf j}}c_{s,{\bf i}}\rangle$ is the
matrix element of the lesser Green's function of the scattering
region. Here ${\bf J}_{\sigma{\bf i},\sigma'{\bf j}}$ is the current
from site i to j. Under dc bias, the current is time-independent.
After taking Fourier transform, current ${\bf J}_{\sigma{\bf
i},\sigma'{\bf j}}$ can be written as:
\begin{eqnarray}
{\bf J}_{\sigma{\bf i},\sigma'{\bf j}} =\frac{2e}{\hbar}\int
\frac{dE}{2\pi}~~{\rm Re}\left[ {\bf
G}^<_{\sigma\mathbf{i},\sigma'\mathbf{j}}(E){\bf
H}_{\sigma'\mathbf{j},\sigma\mathbf{i}}\right]\label{DOC0}
\end{eqnarray}

Due to the current conservation, ${\bf J}_{\sigma,{\bf i}}=0$ at
each site inside the scattering region. From now on, we will
calculate the current density from ${\bf J}_{\sigma{\bf
i},\sigma'{\bf j}}$. For the square lattice, it is easy to calculate
the current density by summing over all projections of ${\bf
J}_{\sigma{\bf i},\sigma'{\bf j}}$ along x and y directions as done
in Ref.\onlinecite{JiangHua,lens,LiJian1}. However, it is much more
complicated for graphene because in the tight-binding representation
on graphene, the current can flow from site i to its nearest and
next nearest neighbor sites j. Hence we will use another simple
definition\cite{CurrentDensity} of current density $j = \rho v$
where $\rho$ is the charge density and $v$ is the velocity given by
${\bf v}=-i/\hbar({\bf r}{\bf H}-{\bf H}{\bf r})$. The current
density is
\begin{eqnarray}
{\bf J}_{x/y}({\bf i}) = -\frac{e}{\hbar}\int \frac{dE}{2\pi}~~{\rm
Re}\sum_\sigma\left[{\bf G}^<({\bf r}_{x/y}{\bf H}-{\bf H}{\bf
r}_{x/y})\right]_{\sigma\mathbf{i},\sigma{\bf i}}\label{DOC1}
\end{eqnarray}
where ${\bf r}$ is a diagonal matrix in the discrete real space. It
should be noted that Eq.(\ref{DOC1}) is valid when site i is not on
the interface between the lead and the scattering region. For these
boundary sites, the current density perpendicular to the interface
obtained from Eq.(\ref{DOC1}) should be multiplied by two. This is
due to the following reason. When we have a finite scattering, the
current is conserved if the self-energy is taken into account.
However, for these boundary sites the current from the lead is not
accounted for by Eq.(\ref{DOC1}).

From the Keldysh equation, the lesser Green's function is related to
the retarded and advanced Green's functions,
\begin{eqnarray}
{\bf G}^<(E)={\bf
G}^r(E)\left[\sum_\alpha{\bf\Sigma}^<_\alpha(E)\right]{\bf
G}^a(E)\label{Gl}
\end{eqnarray}
where the sum index $\alpha=L,R$ denote the left and right
semi-infinite lead, ${\bf
\Sigma}^<_\alpha(E)=i{\bf\Gamma}_\alpha(E)f_\alpha(E)$ in
Eq.(\ref{Gl}) is the lesser self energy of the lead-$\alpha$ with
$f_\alpha(E)=f_0(E-eV_\alpha)$ the Fermi distribution function. Here
${\bf\Gamma}_\alpha(E)=i(\Sigma^r_\alpha-\Sigma^a_\alpha)$ and
${\bf\Sigma}^{a/r}_\alpha$ is related to the surface Green's
function which can be calculated using a transfer matrix
method.\cite{transfer} $V_\alpha$ is the external bias in the
terminal-$\alpha$. In general, $G^<(E)$ can be divided into
equilibrium and non-equilibrium parts,\cite{lens}
\begin{eqnarray}
{\bf G}^<(E)&=&{\bf G}^r(E)\left[if_0(E)\sum_\alpha{\bf\Gamma}_\alpha(E)\right]{\bf G}^a(E)\nonumber \\
&+& {\bf
G}^r(E)\left[i\sum_\alpha\left\{f_\alpha(E)-f_0(E)\right\}{\bf\Gamma}_\alpha(E)\right]{\bf
G}^a(E)
\end{eqnarray}
where the equilibrium term can only generate persistent
current\cite{foot} and does not contribute to the
transport,\cite{persistent} so it can be dropped out from now on. It
is the non-equilibrium term that gives the response to the electron
injection from the source lead. Setting source bias $V_s=-V$ and
drain bias $V_d=0$, we have
\begin{eqnarray}
{\bf G}^<(E)=i{\bf
G}^r(E)\left[f_s(E)-f_0(E)\right]{\bf\Gamma}_s{\bf
G}^a(E)\label{Gless}
\end{eqnarray}
where ${\bf \Gamma}_s$ is the linewidth function of source lead.
Substituting Eq.(\ref{Gless}) into Eq.(\ref{DOC1}), the differential
local current density vector $d{\bf J}_{x/y}/dV$ at site ${\bf i}$
can be expressed in the following form:\cite{foot1}
\begin{eqnarray}
d{\bf J}_{x/y}(\mathbf{i})/dV 
&=& e\sum_{\sigma}{\rm Re}\left[\rho{\bf
v}_{c,x/y}\right]_{\sigma{\bf i},\sigma{\bf i}}\label{DOC2}
\end{eqnarray}
with
\begin{eqnarray}
&&\rho=\frac{1}{2\pi}{\bf G}^r(E_F+eV){\bf
\Gamma}_s(E_F+eV){\bf G}^a(E_F+eV)\label{rho} \\
&&{\bf v}_{c,x/y}=-\frac{i}{\hbar}({\bf r}_{x/y}{\bf H}-{\bf H}{\bf
r}_{x/y})\nonumber
\end{eqnarray}
where $\rho$ and ${\bf v}_{c,x/y}$ are density matrix with incident
energy $E_F+eV$ and velocity matrix in the central scattering
region, respectively and we have assumed the temperature is zero.

XXX

When the electron is in the eigenmodes of the semi-infinite lead
there is no scattering, the the linewidth function ${\bf \Gamma}_s$
of source lead is related to the incident velocity in the form
$\hbar{\bf v_s}={\bf U^\dagger \Gamma_s U=\tilde{\bf
\Gamma}_s}$\cite{XiaKe} where ${\bf v}_s$ is diagonal matrix
composed by the nonzero velocity in propagating mode and zero
velocity in evanescent mode, ${\bf U}$ is ranked by eigenmodes
including propagating mode and evanescent mode, here ${\bf U}$ can
be considered as a unitary transformation matrix which transforms
the general Hilbert space into the eigen channel space of lead.
Then, $\tilde{\bf \Gamma}_s$ can be regarded as incoming velocity
matrix in source lead with relation ${\bf v}_s=(1/\hbar)\tilde{\bf
\Gamma}_s$ [for detail please refer to Ref.\onlinecite{XiaKe}]. In
the calculation, we can peak up only the propagating mode to
construct the effective linewidth function $\tilde{\bf \Gamma}_s$ as
\begin{eqnarray}
\bar{\bf \Gamma}_s=\sum_n{\bf \Lambda}_n{\bf v}_{s,n}{\bf
\Lambda}^\dagger_n
\end{eqnarray}
where the sum is taken over propagating modes. ${\bf \Lambda}_n$ is
the $n$-th column of matrix $[{\bf U}^\dagger]^{-1}$ which is
related to the $n$-th propagating mode. Then from
Eq.(\ref{DOC2}),(\ref{rho}), we can write the differential local
current density vector in $n$-th propagating eigenchannel:
\begin{eqnarray}
d{\bf J}_{n,x/y}(\mathbf{i})/dV = \frac{e}{2}\sum_{\sigma}\left[{\rm
Re}(\rho_n{\bf v}_{c,x/y})\right]_{\sigma{\bf i},\sigma{\bf i}}
\label{DOC3}
\end{eqnarray}
where
\begin{eqnarray}
\rho_n=\frac{1}{2\pi}{\bf G}^r{\bf \Lambda}_{s,n}{\bf v}_{s,n}{\bf
\Lambda}^\dagger_{s,n}{\bf G}^a\label{rhon}
\end{eqnarray}

Concerning for the scalar current flowing into the drain lead $J_d$
and the conductance $G=dJ_d/dV$, we can replace ${\bf v}_{c,x/y}$
with ${\bf v}_d=(1/\hbar)\tilde{\bf \Gamma}_d$ in Eq.(\ref{DOC2})
where ${\bf \Gamma}_d$ is the effective outgoing velocity matrix of
drain lead similar as effective incident velocity matrix ${\bf
v}_s=(1/\hbar)\tilde{\bf \Gamma}_s$ in source lead. Then,
considering the representation transformation, from Eq.(\ref{DOC2}),
(\ref{rho}), we can get the Landauer-Buttikker
formula\cite{Landuaer}, which leads to conductance in zero
temprature
\begin{equation}
G(E_F+eV) = \frac{e}{\hbar}{\rm Tr}\left[{\rm Re}(U\rho
U^\dagger\tilde{\bf \Gamma}_{d})\right]=\frac{e}{h} T\label{cond}
\end{equation}
where $T={\rm Tr}\left[{\rm Re}({\bf G}^r{\bf \Gamma}_s{\bf G}^a{\bf
\Gamma}_{d})\right]$ is the transmission coefficient from source
lead to drain lead.

\section{numerical results and discussion}

In the numerical calculations, the energy is measured in the unit of
the nearest neighbor coupling constant $t$. For the first model,
$t=B/a^2$ with the square lattice constant $a=5nm$. For the second
and third model in the honeycomb lattice, $t=2\hbar v_F/(3a)$ with
the carbon-carbon distance $a=0.142nm$ and the Fermi velocity
$v_F=0.89\times10^6ms^{-1}$ as in a real graphene sample.\cite{ref3}
The size of the scattering region [the green (or gray in print)
region] $M \times N$ is described by integers $M$ and $N$
corresponding to the width and length, respectively. For example, in
Fig.1, the width $W=M a$ with $M=4$, the length $L=N a$ with $M=7$
in the panel (a), and the width $W=M \times 3a$ with $M=3$, the
length $L=N\times \sqrt{3}b$ with $M=7$ in the panel (b).

\begin{figure}
\includegraphics[bb=1mm 140mm 205mm 229mm,
width=8.5cm,totalheight=4.0cm, clip=]{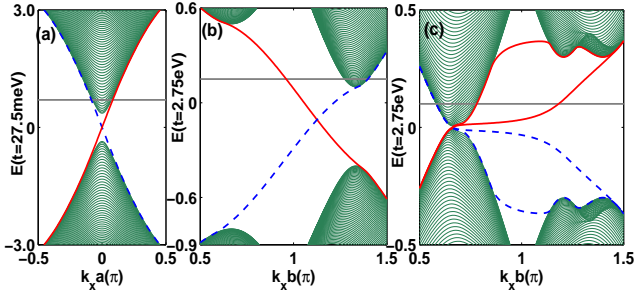} \caption{
(Color online) The band structure for first [panel(a)], second
[panel(b)] and third [panel(c)] model. The horizontal gray lines
marked the Fermi energy $E_F=0.7,0.15$ and $0.1$ in three models,
respectively. The blue (dark gray in print) dashed lines denote the
edge state located in the down edge, and the red (light gray) lines
is for the edge state located in the up edge.}
\end{figure}

In all three models Eq.(\ref{Diracr}), (\ref{Haldaner}) and
(\ref{Qiaor}), edge state exists in the absence of external magnetic
field, which induce the quantum anomalous Hall effect. In Fig.2, we
plot the band structure (for calculation see Ref.\onlinecite{band})
and indicate the Fermi energy (the gray lines) in three models. We
can see the unidirectional edge states along the each edge, down
edge [the state in the blue (or dark gray in print) dashed lines] or
up edge [the state in the red (or light gray in print) lines] in
these models. For the first model [panel(a)], since e-h symmetry is
kept so that we can focus only on electrons, i.e., only the positive
energy is considered. This is different from the HgTe/CdTe quantum
well, in which e-h symmetry is broken and the edge state is more
localized along the edge for the positive energy than the negative
energy. In the calculation we set Fermi energy $E_F=0.7t$,
$A/2a=1.35t$, $C=3.65t$, $B/a^2=t=27.5meV$, $a=5nm$. For the second
model [panel (b)], the following parameters are used: Fermi energy
$E_F=0.15t$, the nearest neighbor coupling constant $t=27.5eV$ that
is used as the energy unit in the graphene models, the next-nearest
neighbor coupling strength $t_n=0.1t*e^{i\pi/3}$, the staggered
sublattice potential $t_o=0.2t$. In this model, e-h symmetry is
broken by the next-nearest neighbor coupling $t_n$, and the edge
states are favored the hole system (negative energy). In the third
model, different from the second one, the next-nearest neighbor
coupling is absent, the Rashba SOC $t_r=0.18t$ and exchange energy
$t_e=0.2t$ are considered, Fermi energy $E_F=0.1t$. Due to the
Rashba SOC, the spin degeneracy is lifted and there are two
unidirectional edge states [red
or blue
lines in panel (c)] corresponding to different spins. In the second
and third models, we find that the staggered sublattice potential
$t_o$ is important to study the nature of TAI phenomena. First,
$t_o$ breaks the inversion symmetry, the left [red
line in Fig.2(b), blue
line in Fig.2(c)] and right flowing edge states as well as the bulk
states are asymmetrically distributed as shown in Fig.2(b) and (c).
As a result, the left and right injected current density are
distributed differently, although they contribute to the same total
current. Second, $t_o$ makes the bulk gap smaller, which leads to
the coexistence of the robust edge state and bulk states in certain
energy window $\Delta E$. For instance $\Delta E=[0.1,0.5]$ in
Fig.2(b) where only one edge state near the upper edge that coexist
with the bulk states. In Fig.2(c), the bulk gap is closed
completely,
as will be discussed later that for this case it is harder to induce
the TAI phenomena. However this model can give us some insight about
TAI phenomena.

\begin{figure}
\includegraphics[bb=10mm 8mm 212mm 237mm,
width=6.5cm,totalheight=7.0cm, clip=]{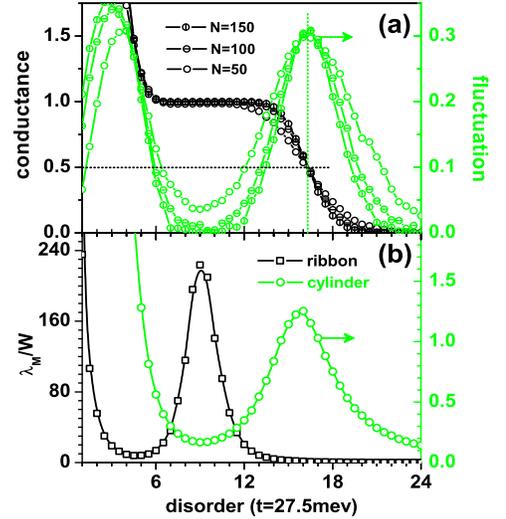} \caption{
(Color online) Conductance, fluctuation and localization length vs
disorder for the first model. Panel (a): conductance [black lines]
and fluctuation [green or gray in print lines] in a square lattice
with different widths $W=M \times a$. Panel (b): renormalized
localization length of a 2D strip sample [black line] and cylinder
sample [green or gray in print line], the width of strip or the
diameter of cylinder are all 50a. }
\end{figure}
\begin{figure}
\includegraphics[bb=11mm 10mm 206mm 237mm,
width=6.5cm,totalheight=7.0cm, clip=]{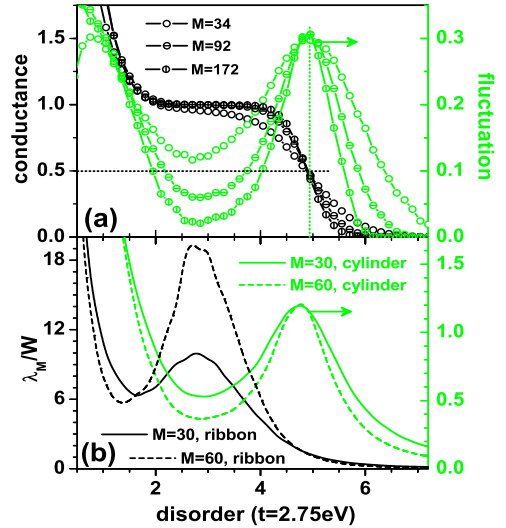}
\caption{(Color online) The same as Fig.3 except that it is for the
second model.}
\end{figure}

In Fig.3(a) and Fig.4(a), we plot conductance and its fluctuation vs
disorder for the first model in a square lattice and the second
model in a honeycomb lattice, respectively. In Fig.3(a), Fig.4(a)
and Fig.5, all data are obtained by averaging over 5000
configurations. In Fig.3(b) and Fig.4(b), we also plot the
renormalized localization length $\xi_M/W$ of a 2D strip with the
width $W=Ma$ for the square lattice [Fig.3(b)] and $W=M \times 3a$
for the honeycomb lattice [Fig.4(b)]. Furthermore, in order to
highlight the topological nature, we also plot the renormalized
localization length $\xi_M/W$ in a 2D cylinder sample (where the
edge state is removed) with diameter $W$. Here the Fermi energy
$E_F$ is set in the bulk state [see Fig.2]. In the clean system with
$w=0$, the conductance is contributed by bulk states and one edge
state. When disorder is introduced, the following observations are
in order: (1) the small disorder rapidly suppresses the conductance
and enhances its fluctuation. This means that the system is leading
to the diffusive regime, in which the bulk extended state is
scattered by impurities and gradually becomes localized by disorder.
As a result, the localization length decreases accordingly for both
strip and cylinder samples. (2) at moderate disorders, in the strip
sample, the conductance stops decreasing and develops into a
quantized plateau, a new phase occurs. At the same time, the
fluctuation is deduced to almost zero, indicating the formation of
either a bulk insulator or metallic phase. The abruptly increased
localization length (black lines) in panel (b) clearly indicates the
metallic phase induced by edge state. For the cylinder sample,
however, because of the absence of the edge state, the system can
only develop into the bulk insulator, so the localization length
continues to decrease. Our results show that the wider the ribbon
is, the larger $\xi_M/W$ in ribbon geometry and the smaller
$\xi_M/W$ in cylinder geometry [see Fig.4(b)]. Therefore, concerning
$\xi_M/W$, the peak structure in strip geometry and the valley
structure in cylinder geometry clearly indicate the topological
nature of the TAI phenomena. (3) when the disorder is strong, the
quantized conductance plateau in the strip sample starts to
deteriorate and its fluctuation begins to increase. This shows that
the edge state will be destroyed at the strong disorder and the
system will enter the insulating regime.

For the cylinder sample, there is only one system transforms from
band insulator to Anderson insulator. The peak of $\xi_M/W$
indicates the transition points.
(4) at the critical disorder strength $w_c$ MIT occurs. $w_c$ is
marked in Fig.(3) and (4) by the vertical green
gray in print) dotted lines, where conductance $G=0.5\frac{e^2}{h}$
and $\xi_M/W\simeq$. For cylinder sample, $\xi_{M\rightarrow
\infty}$ diverges at $w_c$. It should be noted that despite of the
different topological nature, the MIT occurs at nearly the same
$w_c$ for strip and cylinder samples. (5) in the quantized
conductance regime, increasing the sample size, the quantized
plateau is much wider with smaller fluctuation. This is not
surprising since for the larger sample, the bulk state is scattered
more frequently while the overlap of opposite unidirectional edge
states is smaller. This size effect is clearly seen in Fig.5, in
which the quantized conductance and its fluctuation in the midpoint
[$w=9$ for panel (a) and $w=2.25$ for panel(b)] of conductance
plateaus are plotted vs width of square sample for the fist model
and second models. We can see with the increasing of the sample
size, the conductance eventually saturates to the quantized value,
its fluctuation, however, continues to drop. It is expected that
when the size is large enough, the complete quantized plateau will
be formed with no conductance fluctuation.

\begin{figure}
\includegraphics[bb=11mm 10mm 213mm 227mm,
width=6.0cm,totalheight=7.0cm, clip=]{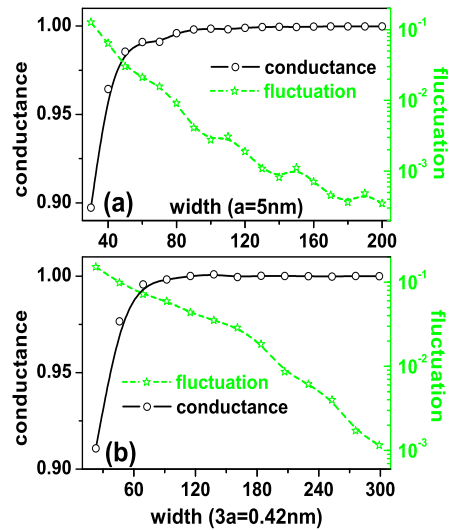}
\caption{(Color online) Conductance [black lines] and fluctuation
[green or gray in print lines] of a 2D square sample vs sample size
in the first model for the fixed disorder strength $w=9$ [panel (a)]
and in the second model for the fixed disorder strength $w=2.25$
[panel (b)]. }
\end{figure}

Up to now, we have confirmed that in addition to the HgTe/CdTe
quantum well, the TAI, i.e., disorder induce quantized conductance
can also occur in other systems such as the Dirac model and the
graphene system.
XXX Since the phenomenon of TAI is not the
particular properties of HgTe/CdTe quantum well, there must be the
common nature of TAI phenomena in different systems. For this
reason, we study the scattering process in different systems through
monitoring the local current density.

\begin{figure}
\includegraphics[bb=2mm 51mm 194mm 245mm,
width=8.5cm,totalheight=8.0cm, clip=]{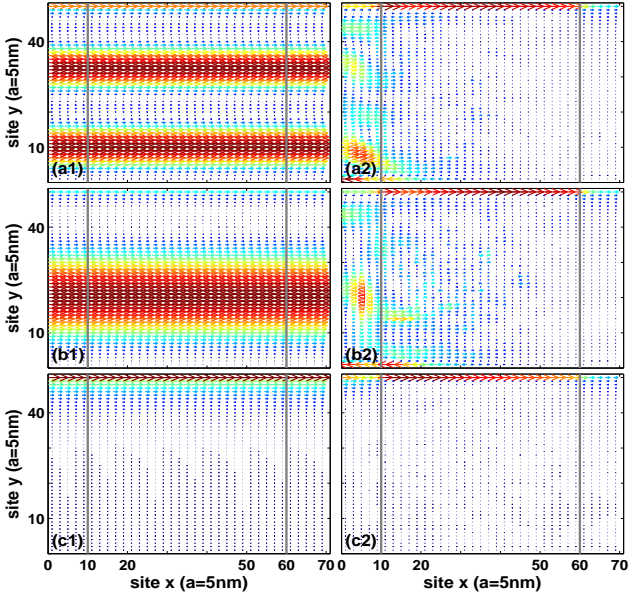}
\caption{(Color online) Left injected local differential current
density vector distribution for the lowest three sub-channels in the
first model. The direction and the length of arrow in every site
show the direction and the magnitude of current density vector in
this site. Concerning a visual effect, the magnitude of current
vector is also expressed by different color. With the increasing
magnitude, the color changes from blue to red. The left and right
column are corresponding to the clear sample and dirty sample with
disorder strength $w=6$.}
\end{figure}

First we study the Dirac model. Due to the inversion symmetry in the
Dirac model, the left and right injected current are equivalent,
here we focus only on the left injected current. We have set Fermi
energy to be $E_F=0.7$ for Fig.6. In Fig.6, Fig.\ref{DensHa1},
Fig.\ref{DensHa1}, and Fig.\ref{DensQiao}, all data are obtained by
averaging over 1000 configurations. For $E_F=0.7$, there are several
transmission channels and we have studied the first three
transmission channels, one is edge state and the other two are bulk
states. In Fig.6(a)-(c), we plot the left injected local
differential current density distribution from the lowest three
transmission-channel including the edge channel [panel (c)] and the
first [panel (b)] and second [panel (a)] bulk channel in a square
sample [between two vertical gray lines]. For the clean system with
$w=0$ [the left column], the system is ballistic and there is no
scattering. So, the current density from all transmission channels
are distributed along the transport direction. For the left column
of Fig.6, we can see the right going edge state on the upper edge
and the extended bulk state with one or two transverse peaks in the
current density. Note that the coming eigen-channels are classified
according to the lead. When disorder is present, due to the mode
mixing the electron in edge channel can be scattered into bulk
channels and vice versa. As a result, the eigen-channels of the lead
are no longer the eigen-channels of the system. Corresponding to the
left column of Fig.6, in right column, we plot the current density
distribution at the entrance of the TAI (quantized conductance) with
$w=6$. We can see that the right going edge current in all
transmission-channels of the lead are well protected.

\begin{figure}
\includegraphics[bb=9mm 9mm 192mm 149mm,
width=8.5cm,totalheight=6.5cm, clip=]{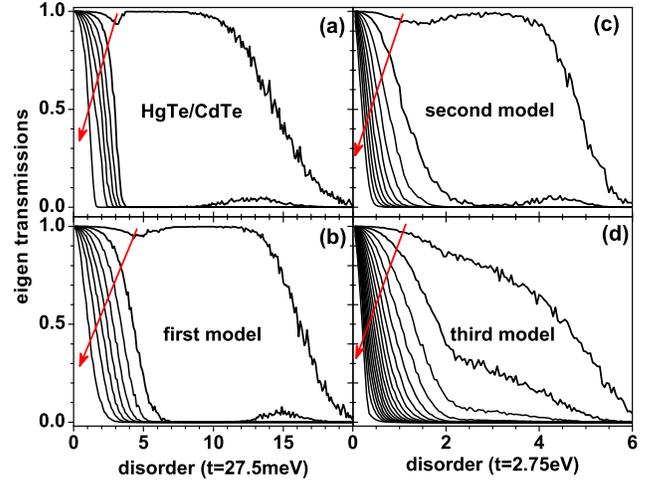} \caption{(Color
online) transmissions of system eigen channels in a square sample
for the first model [panel (b), the width $W=100a$], the second
model [panel (c), the width $W=92*3a$], the third model [panel (d),
the width $W=100*3a$] and HgTe/CdTe quantum well [panel (a), the
width $W=100a$]. Different lines long the red arrow correspond to
the eigen edge channel, the first, second, ..., eigen bulk channel.
}\label{trans}
\end{figure}

In order to study the nature of the edge state in the presence of
disorders, we have calculated the eigen-spectrum of the transmission
matrix $\Gamma_L G^r \Gamma_R G^a$.\cite{Qiao_new} In
Fig.\ref{trans}(b), we plot all the transmissions of eigen channels
for the the first and second model. For comparing, in Fig.7(a), we
also plot the eigen transmissions of HgTe/CdTe quantum well in which
except for the terms in the first model, the term of $D{\bf
k}\cdot{\bf k}$ are also included. Here, we set $D/a^2=0.75t$ with
$t=B/a^2$ and the other parameters are same as in the first model.
In Fig.\ref{trans}, all data are obtained by averaging over 100
configurations. We found that there is always one eigen-value that
is nearly equal to one which is identified as edge state by
examining its current density profile. It illuminate that due to the
topological nature, the edge state is always protected during the
transport, although it may be scattering into the bulk channels of
lead.

From Fig.\ref{trans}, we also found that the eigen transmissions of
the first and second model in this paper exhibit the same properties
as that in the HgTe/CdTe quantum well. It strongly implies they have
the same

\begin{figure}
\includegraphics[bb=2mm 111mm 194mm 245mm,
width=8.5cm,totalheight=5.5cm, clip=]{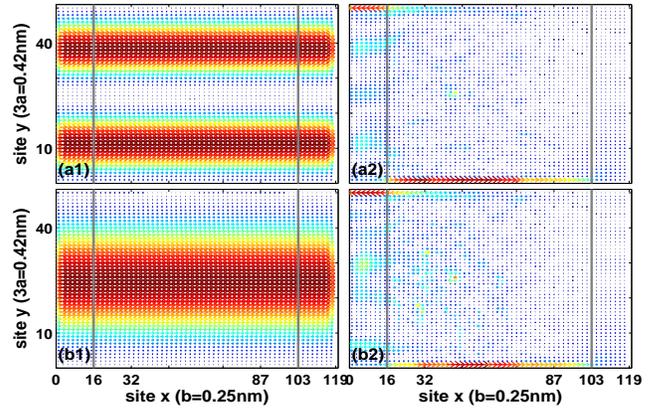}
\caption{(Color online) Panel (a-b): Left injected local
differential current density vector distribution for the lowest two
sub-channels in the second model. The left and right column are
corresponding to the clear sample and dirty sample with disorder
strength $w=1.8$.} \label{DensHa1}
\end{figure}
\begin{figure}
\includegraphics[bb=2mm 51mm 194mm 245mm,
width=8.5cm,totalheight=8.0cm, clip=]{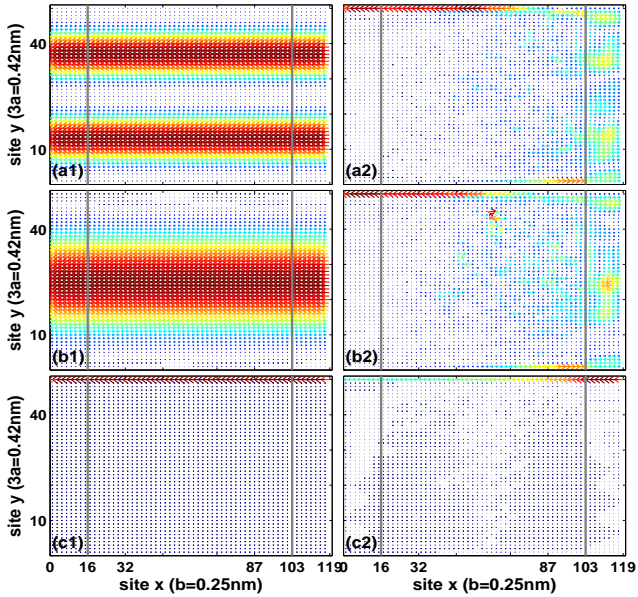}
\caption{(Color online) Panel (a-c): Right injected local
differential current density vector distribution for the lowest
three sub-channels in the second model. The left and right column
are corresponding to the clear sample and dirty sample with disorder
strength $w=1.8$.} \label{DensHa2}
\end{figure}

Next, we examine the graphene model. We set $E_F=0.15$ for
Fig.\ref{DensHa1} and Fig.\ref{DensHa2}. In the graphene model, the
staggered sublattice potential is considered, which breaks the
inversion symmetry, and the configuration of left and right injected
current density are different. In a clean system, the left (right)
injected current is contributed by the states that have a positive
(negative) group velocity, i.e., $\partial E/\partial k_x>0(<0)$.
For the fixed Fermi energy, the crossing of Fermi energy and energy
band determines the momentum $k$ of all propagating states. From the
band structure of the graphene model [Fig.2(b) and (c)], we can see
that the edge state is absent for the right propagating band in the
second model while in the third model, it is absent for the left
propagating state.

In Fig.\ref{DensHa1} and Fig.\ref{DensHa2}, we plot the left and
right injected current density in the second model, respectively. We
can see that for the left injected current, although edge current is
absent in clean limit, it emerges when disorder is turned on. This
can be understood as follows.\cite{Beenakker} When the disorder is
present, the effective medium theory similar to the one in
Ref.\onlinecite{Beenakker} shows that the parameters in the second
model such as $t_n$ and $E_F$ are all renormalized and depend on
disorder strength. After renormalization due to the disorder, the
new Fermi level (renormalized) is inside the bulk gap of the new
band structure giving rise to a true edge state.

Besides, the random disorder also eliminates the difference between
states with different momentum $k$ due to its random distribution.
So, although the profiles of left and right injected current density
are different in the clean system, when the moderate disorder is
considered they tend to be the same since the bulk state are killed
and only edge state survives [see Fig.7 and Fig.8]. Of course, the
left going and right going edge states are along the opposite edge.
Similar to the first model, in clean limit, the current density in
all eigenchannels are kept and uniformly distributed along the
transport direction, except they have different profiles for left
and right injected current density.

XXX

Form Fig.6, \ref{DensHa1} and \ref{DensHa2}, it seems that the
phenomena of disorder induce quantized conductance can appears in
any system where quantum anomalous Hall effect is hold. However, it
shoulder be noted that in the first and second model, the edge state
is robust so that they can penetrate into the scattering region
under the interface scattering. In this way, the quantized is
formed. In the following we will study another quantum anomalous
Hall system, it is the third model. Similar to the second model, due
to the inversion asymmetry, the configuration of the left and right
injected current density are different in the clean limit, here we
focus only on the left injected case in which the edge current is
present in clean system. In the third model, the bulk energy gap is
closed by the staggered sublattice potential, there is no pure edge
state for any incident energy, so the edge state is not so robust as
in the first and second model. Furthermore, due to inversion
asymmetry, one of the two unidirectional edge states turns into bulk
state [see Fig.2(c)]. So the disorder induced conductance must be
$e^2/h$ if any.

As a result, because the edge state is not so robust, the back
following edge current is strong, although there is one unambiguous
unidirectional edge state [see left column of Fig.\ref{DensQiao}],
it is drastically reflected by the interface scattering, then the
edge current can't deeply penetrate into the scattering region when
moderate disorder strength is considered, which can be clearly seen
in Fig.\ref{DensQiao}(c2). On the other hand, the edge state that
enter into the scattering region can always survive in a moderate
disorder [see Fig.\ref{DensQiao}(a2) and (b2)]. As a result, the
conductance can't be quantized but flattened by the moderate
disorder, which is different from the first and second model, but
they share the same nature, that can also be seen from
Fig.\ref{trans}(c) and (d).

In Fig.\ref{trans}(c) and (d) we plot the system eigen transmissions
of the second and third model. We can see for all model (the first,
second, third model and the HgTe/CdTe quantum well), the
transmissions of eigen bulk state are rapidly reduced to zero in a
weak disorder, while the eigen edge state can be kept in the
moderate disorder. When the eigen edge state is robust, it can be
completely kept [transmission equal to one, see panel (a), (b) and
(c)], but when the eigen edge state is not so robust, it can only be
partly kept [transmission is smaller than one, see panel (d)]. So we
can conclude that in the system where robust edge state and bulk
state coexist, the edge state is maintained while the bulk states
are killed in the moderate disorder, which leads to the flat
conductance. If the edge state is robust enough to resist the
disorder that is so strong that the bulk state are completely
killed, quantized conductance can be formed and TAI phenomena
appears.

\begin{figure}
\includegraphics[bb=2mm 51mm 194mm 245mm,
width=8.5cm,totalheight=8.0cm, clip=]{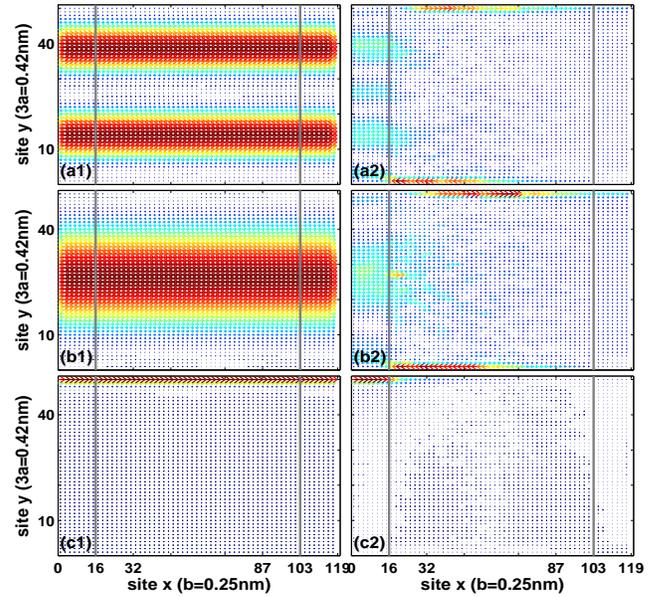} \caption{(Color
online) Panel (a-c): Left injected local differential current
density vector distribution for the lowest three sub-channels in the
third model. The left and right column are corresponding to the
clear sample and dirty sample with disorder strength $w=1.8$.}
\label{DensQiao}
\end{figure}

\bigskip

\section{conclusion}
In summary, the disorder effect is studied in several quantum
anomalous Hall system. When beyond the energy gap, the edge state
and bulk state often coexist in clean system or weak disorder
system. In a moderate disorder, the inter-band scattering happens.
If the edge state is robust enough to resist the interface
scattering, it can penetrate deep into scattering region where the
edge state and bulk state coexist in all of the coming eigen
channels due to the mod mixing. Due to the topological nature, the
bulk states are drastically killed and edge state survives. In the
eigen spectrum of the whole system, the eigen channel corresponding
to the edge state is kept well and the other bulk channels are
completely killed, which leads to a flattened conductance
accompanied with the rapidly decreased fluctuation. In several
spatial systems such as HgTe/CdTe quantum well, modified Dirac model
and the graphene system with next-nearest coupling and staggered
sublattice potential, the edge state is very robust so that it can
survive in a disorder that is so strong that the bulk states are
completely killed by them. Without the bulk states, the
unidirectional edge state located in up or down edge can't be
transformed to the opposite edge, the the back following current is
then prohibited. As a result, the quantized conductance can be
formed and TAI phenomena that disorder induce quantized conductance
appears.

$${\bf ACKNOWLEDGMENTS}$$
We gratefully acknowledge the financial support by a RGC grant (HKU
705409P) from the Government of HKSAR.


\begin{thebibliography}{00}
\bibitem{foot}
Here the persistent current is due to the magnetic vector potential
$A$. If $A=0$, equilibrium current is zero because the left and
right lead contribute opposite current density considering the
symmetry properties.\cite{CurrentDensity}
\bibitem{foot1}
The expression should be modified when the site is on the boundary
of the scattering region.
\end{thebibliography}
\end{document}